\let\includefigures=\iffalse
%
\let\useblackboard=\iftrue
%
%
\newfam\black
\input harvmac
\includefigures
\message{If you do not have epsf.tex (to include figures),}
\message{change the option at the top of the tex file.}
\input epsf
\def\figin{\epsfcheck\figin}\def\figins{\epsfcheck\figins}
\def\epsfcheck{\ifx\epsfbox\UnDeFiNeD
\message{(NO epsf.tex, FIGURES WILL BE IGNORED)}
\gdef\figin##1{\vskip2in}\gdef\figins##1{\hskip.5in}
\else\message{(FIGURES WILL BE INCLUDED)}%
\gdef\figin##1{##1}\gdef\figins##1{##1}\fi}
\def\DefWarn#1{}
\def\figinsert{\goodbreak\midinsert}
\def\ifig#1#2#3{\DefWarn#1\xdef#1{fig.~\the\figno}
\writedef{#1\leftbracket fig.\noexpand~\the\figno}%
\figinsert\figin{\centerline{#3}}\medskip\centerline{\vbox{\baselineskip12pt
\advance\hsize by -1truein\noindent\footnotefont{\bf Fig.~\the\figno:} #2}}
\bigskip\endinsert\global\advance\figno by1}
\else
\def\ifig#1#2#3{\xdef#1{fig.~\the\figno}
\writedef{#1\leftbracket fig.\noexpand~\the\figno}%
\global\advance\figno by1}
\fi
\useblackboard
\message{If you do not have msbm (blackboard bold) fonts,}
\message{change the option at the top of the tex file.}
\font\blackboard=msbm10 scaled \magstep1
\font\blackboards=msbm7
\font\blackboardss=msbm5
\textfont\black=\blackboard
\scriptfont\black=\blackboards
\scriptscriptfont\black=\blackboardss

\else

\fi
%
\def\yboxit#1#2{\vbox{\hrule height #1 \hbox{\vrule width #1
\vbox{#2}\vrule width #1 }\hrule height #1 }}
\def\fillbox#1{\hbox to #1{\vbox to #1{\vfil}\hfil}}
\def\ybox{{\lower 1.3pt \yboxit{0.4pt}{\fillbox{8pt}}\hskip-0.2pt}}

\def\comments#1{}

\def\p{\partial}

\def\CM{{\cal M}}

\def\nl{\hfill\break}

\def\II{\relax{I\kern-.10em I}}
\def\IIa{{\II}a}

\def\IZ{\relax\ifmmode\mathchoice
{\hbox{\cmss Z\kern-.4em Z}}{\hbox{\cmss Z\kern-.4em Z}}
{\lower.9pt\hbox{\cmsss Z\kern-.4em Z}}
{\lower1.2pt\hbox{\cmsss Z\kern-.4em Z}}\else{\cmss Z\kern-.4em
Z}\fi}
\def\IB{\relax{\rm I\kern-.18em B}}
\def\IC{{\relax\hbox{$\inbar\kern-.3em{\rm C}$}}}
\def\ID{\relax{\rm I\kern-.18em D}}
\def\IE{\relax{\rm I\kern-.18em E}}
\def\IF{\relax{\rm I\kern-.18em F}}
\def\IG{\relax\hbox{$\inbar\kern-.3em{\rm G}$}}
\def\IGa{\relax\hbox{${\rm I}\kern-.18em\Gamma$}}
\def\IH{\relax{\rm I\kern-.18em H}}
\def\II{\relax{\rm I\kern-.18em I}}
\def\IK{\relax{\rm I\kern-.18em K}}
\def\IP{\relax{\rm I\kern-.18em P}}

%

\def\inbar{\,\vrule height1.5ex width.4pt depth0pt}

\def\p{\partial}
\def\bp{{\bar \p}}

\font\cmss=cmss10 \font\cmsss=cmss10 at 7pt
\def\IR{\relax{\rm I\kern-.18em R}}

\def\BC{\IC}

\def\lp10{l_P^{10}}
\def\lp11{l_P^{11}}
\def\R11{R_{11}}

\def\wM{{\widetilde{M}}}
\def\wm{{\widetilde{m}}}
\def\wg{{\widetilde{g}}}
\def\wl{{\widetilde{l}}}

\def\wR{{\widetilde{R}}}
\def\wsig{{\widetilde{\sigma}}}
\Title{\vbox{\baselineskip12pt\hbox{hep-th/9710178}
\hbox{LBNL-40889}
\hbox{RU-97-85}\hbox{UCB-PTH-97/51}}}
{\vbox{
\centerline{Why Matrix Theory is Hard} }} 
\centerline{Michael R. Douglas$^{1,2}$ and Hirosi Ooguri$^{3,4}$}
\bigskip
\centerline{$^1$ Institut des Hautes \'Etudes Scientifiques}
\centerline{Le Bois-Marie, Bures-sur-Yvette, 91440 France}
\medskip
\centerline{$^2$ Department of Physics and Astronomy}
\centerline{Rutgers University, Piscataway, NJ 08855--0849}
\medskip
\centerline{$^3$ 366 Le\thinspace Conte Hall, Department of Physics}
\centerline{University of California, Berkeley, CA 94720-7300}
\medskip
\centerline{$^4$ Theory Group, Mail Stop 50A-5101, Physics Division}
\centerline{Lawrence Berkeley National Laboratory, Berkeley, CA 94720}
\medskip
\centerline{\tt douglas@ihes.fr, ooguri@thsrv.lbl.gov}
\bigskip
Recently Sen and Seiberg gave a prescription for constructing the
matrix theory in any superstring background. We use their
prescription to test the finite $N$ matrix theory conjecture
on an ALE space. Based on our earlier work with Shenker, we find
a sharper discrepancy between matrix theory computation and
supergravity prediction. We discuss subtleties in the light-front
quantization which may lead to a resolution to the discrepancy.

\Date{October 1997}
%
\nref\BFSS{T. Banks, W. Fischler, S. H. Shenker and L. Susskind, ``M
Theory as a Matrix Model: a Conjecture,''
Phys. Rev. D55 (1997) 5112-5128; hep-th/9610043.}
\lref\strings{M. R. Douglas, ``D-branes and Matrix Theory in Curved
Space,'' talk given at Strings '97; hep-th/9707228.}
\lref\DHN{B. de Wit, J. Hoppe and H. Nicolai,
Nucl.Phys. {\bf B 305 [FS 23]} (1988) 545.}
\lref\BST{E. Bergshoeff, E. Sezgin and P. K. Townsend,
Phys. Lett. 189B (1987) 75;
Ann. Phys. 185 (1988) 330.}
\lref\DLN{B. de Wit, M. L\"uscher and H. Nicolai,
Nucl.Phys. {\bf B 305 [FS 23]} (1988) 545.}
\lref\dvv{R. Dijkgraaf, E. Verlinde, H. Verlinde, ``Matrix String Theory,''
Nucl. Phys. B500 (1997) 43-61; hep-th/9703030.}
\lref\ggv{M. B. Green, M. Gutperle, P. Vanhove, ``One Loop in Eleven
Dimensions,'' hep-th/9706175.}
\lref\DLP{J.~Dai, R.~G.~Leigh and J.~Polchinski,
Mod. Phys. Lett. {\bf A4} (1989) 2073.}
\lref\DKPS{M. R. Douglas, D. Kabat, P. Pouliot and S. Shenker,
``D-branes and Short Distances in String Theory,''
Nucl. Phys. B485 (1997) 85-127; hep-th/9608024.}
\nref\Banks{T. Banks, ``The State of Matrix Theory,'' talks given
at SUSY '97 and Strings '97, hep-th/9706168.}
\nref\my{T. Maskawa and K. Yamawaki, ``The Problem of $P^+ =0$ mode in
the Null Place Field Theory and Dirac's Method of Quantization,''
Prog. Theor. Phys. 56 (1976) 270.}
\nref\bp{C. Pauli and S. Brodsky, ``Discretized Light Cone
Quantization: Solution to a Field Theory in One Space  One Time
Dimensions,'' Phys. Rev. D32 (1985) 2001.}
\nref\sussLC{L. Susskind, ``Another Conjecture about M(atrix) Theory,''
hep-th/9704080.}
\nref\dos{M. R. Douglas, H. Ooguri and S. H. Shenker,
``Issues in M(atrix) Theory Compactification,''
Phys. Lett. {\bf B402} (1997) 36-42; hep-th/9702203.}
\nref\sen{A. Sen, ``D$0$ Branes on $T^n$ and Matrix Theory,'' hep-th/9709220.}
\nref\sei{N. Seiberg, ``Why is the Matrix Model Correct?'' hep-th/9710009.}
\nref\helpol{S. Hellerman and J. Polchinski, as referred to in \bbpt.}
\nref\malda{J. Maldacena, ``Branes Probing Black Holes,'' talk
presented at Strings 97; hep-th/9709099.}
\nref\bbpt{K. Becker, M. Becker, J. Polchinski and A. Tseytlin,
``Higher Order Graviton Scattering in M(atrix) 
Theory,''
hep-th/9706072.}
\lref\dcurve{M. R. Douglas, ``D-branes in Curved Space,'' to appear in
Adv. Theor. Math. Phys.;
hep-th/9702048.}
\lref\polgeo{J. Polchinski, hep-th/9611050.}
\lref\egs{M. R. Douglas, hep-th/9612126.}
\lref\tseytlin{A. Tseytlin, hep-th/9701125.}
\lref\dg{M. R. Douglas and B. Greene, ``Metrics on D-brane
Orbifolds,''
hep-th/9707214.}
\lref\taylor{W. Taylor, hep-th/9611042.}
\lref\guven{R. Gueven, Phys. Lett. B276 (1992) 49.}
\lref\duff{M. J. Duff, hep-th/9608117.}
\lref\dko{M. R. Douglas, A. Kato and H. Ooguri, ``D-brane Actions on
K\"ahler Manifolds,'' to appear in Adv. Theor. Math. Phys.; hep-th/9708012.}
\lref\gorschw{M. Goroff and J. H. Schwarz, ``$D$-Dimensional Gravity
in the Light-Cone Gauge,'' Phys. Lett. 127B (1983) 61. }
\lref\Weinberg{S. Weinberg, ``Photons and Gravitons in Perturbation
Theory: Derivation of Maxwell's and Einstein's Equations,'' Phys. Rev.
138 (1965) B988; ``Ultraviolet Divergences in Quantum Theories of
Gravitation,'' in General Relativity, an Einstein Centenary Survey,
ed. S. Hawking and W. Israel (1979, Cambridge University Press).}
\lref\dm{M. R. Douglas and G. Moore,
``D-branes, Quivers, and ALE Instantons,'' hep-th/9603167.}
\lref\shenker{S. H. Shenker, ``Another Length Scale in String
Theory?'' hep-th/9509132.}
\lref\bcov{M. Bershadsky, S. Cecotti, H. Ooguri and C. Vafa,
``Holomorphic Anomalies in Topological Field Theories,''
Nucl. Phys. B405 (1993) 279; hep-th/9302103.}
\lref\ov{H. Ooguri and C. Vafa, ``Two-dimensional Black Hole and
Singularity of CY Manifolds,'' Nucl. Phys. B463 (1996) 55;
hep-th/9511164.}
\lref\dd{D.-E. Diaconescu and M. R. Douglas, work in progress.}
\lref\lcrefs{K. G. Wilson, T. Walhout, A. Harindranath, W. M. Zhang,
R. J. Perry and S. Glazek, ``Nonperturbative Light-Front QCD,''
Phys. Rev. D49 (1994) 6720-6766;
hep-th/9401153 \nl
G. McCartor, D. G. Robertson, S. S. Pinsky, ``Vacuum Structure of
Two-Dimensional Gauge Theories on the Light Front,''
Phys. Rev. D56 (1997) 1035-1049; hep-th/9612083 \nl
S. Tsujimaru and K. Yamawaki, ``Zero Mode and Symmetry Breaking on the
Light Front,'' hep-th/9704171.}
\lref\dr{M. Dine and A. Rajaraman, ``Multigraviton Scattering in the
Matrix Model,'' hep-th/9710174.}
%

\newsec{Introduction}

Matrix theory \BFSS\ has by now passed a number of significant tests
in toroidal backgrounds, which preserve maximal supersymmetry \Banks.

Its status in curved backgrounds is less clear.
Although definitions have been proposed, which make contact with existing
results in superstring duality, evidence that they
correctly describe dynamics not
constrained by supersymmetry or BPS arguments is still lacking.
It is also not yet clear to what extent the large $N$ limit
of the original proposal is a necessary ingredient, or whether
finite $N$ gauge theory also has a Matrix theory interpretation,
Discretized Light Cone Quantization (DLCQ \refs{\my,\bp}) 
M theory, as proposed by Susskind \sussLC.
In particular, a result which appears to contradict this (which we will
review in detail shortly) was found in studying compactifications on K3
\dos: the definition obtained using the weak string coupling limit
of D$0$-branes does not reproduce the expected interactions between
gravitons; furthermore it is very difficult to do this using any quantum
mechanics of finitely many degrees of freedom as a definition.

The study of toroidal compactification (as reviewed in \Banks)
has shown that additional
light states can be relevant after compactification and this raised a
possibility that these might modify the conclusions of \dos.
Recently significant light was shed on this issue by Sen \sen\ and
Seiberg \sei\ (see also \refs{\helpol,\malda}), who 
showed that all existing definitions of
Matrix theory including DLCQ
can be obtained by carefully following the BFSS prescription,
starting with theories of D$0$-branes in \IIa\ superstring theory,
boosting to the IMF and taking the limit of {\it weak} string coupling.
Seiberg's argument is that this limit should be thought of as turning the
boosted space-like hidden $S^1$ of \IIa\ superstring theory into
the null $S^1$ of DLCQ M theory.
We find the simplicity of this argument compelling.

This prescription naturally generalizes to any superstring
background and for the large volume limit of K3,
it leads directly to the definition studied
in \dos; we show in section 2 that additional states do not contribute.
In section 3 we verify the claims of \dos\ in the regime
of large K3 volume, large distances and large curvature lengths.
This sharpens the contradiction considerably.

We discuss possible resolutions in section 4.
One possibility is that the computation of \dos\ should
not be compared with classical supergravity results; corrections in DLCQ
M theory are important.  However,
assuming distances and curvature lengths are large compared
to the eleven-dimensional Planck scale, we did not find candidates.

Rather, we will suggest as a possible resolution
the idea that there is a subtlety in the argument of \sei,
which appears in taking the limit in which the space-like circle
becomes null.  Doing this step explicitly would require
integrating out modes of zero longitudinal momentum,
and in the analogous study of light-front gauge theory
it is known that this generally produces modifications of the Hamiltonian.
This picture could be tested by checking that
there exist modified Hamiltonians
which reproduce predictions of DLCQ M theory.

\newsec{M theory in curved space}

Recently a precise proposal for Matrix theory in a wide class of
backgrounds, with any amount of supersymmetry,
was formulated in \refs{\sen,\sei}.
Here we will show that, in the case of compactification
on K3, their proposal leads directly to the definition studied in
\dos.

According to \refs{\sen,\sei}, M theory with Planck scale $m_p$
compactified on a light-like circle of radius
$R$  and momentum $p_- = {N \over R}$
is equivalent to $\wM$ theory, which is \IIa\ string theory with Planck scale
$\wm_p$ compactified on a space-like circle of radius
$\wR_s$ with $N$ D$0$-branes in the limit,
\eqn\limit{\wR_s \rightarrow 0, ~~\wm_p \rightarrow \infty,}
while keeping
\eqn\scaleone{ \wR_s \wm_p^2 = R m_p^2}
finite.
Any transverse length scale $l$ (distance between points,
compactification radius, etc.) is rescaled to the length $\wl$
in the $\wM$ theory defined by
\eqn\scaletwo{ \wl \wm_p = l m_p. }

Let $\wg_s$ and $\wl_s=1/\wm_s$ be the string coupling constant and the string
length of $\wM$ theory.  The limit takes
\eqn\relations{\eqalign{
\wm_s^2 &= \wR_s \wm_p^3 \rightarrow \infty \cr
\wg_s &= (\wR_s \wm_p)^{3/2} \rightarrow 0 \cr
{\wl \over \wl_s} &=  \wg_s^{1/3} l m_p \rightarrow 0
}}
and thus, if we do not make further U dualities,
$\wM$ theory is perturbative string theory in the substringy regime
studied in \DKPS.

Let us review some facts about this regime.  The rescaling \relations\
sends the string tension to infinity and at first sight this might seem
to be the usual low energy limit of string theory.  If this were the
case we could simply rescale the effective action
\eqn\stringact{
S = {1\over \wg_s^2 \wl_s^8}\int d^{10}x\
 \sqrt{g}\left(R + \zeta(3)\wl_s^6 R^4 + c \wg_s^2 \wl_s^6 R^4 + \ldots\right)
}
using \relations\ to obtain the effective action of M theory.
However, attempting this rescaling produces
\eqn\naiveact{
S = {R\over l_p^8}\int d^{10}x\
 \sqrt{g}\left(R +
\zeta(3){l_p^6\over\wg_s^2} R^4 + c l_p^6 R^4 + \ldots\right)
}
and we see that the $O(l_s)$ corrections to the action dominate the
M theory terms.

The resolution of this problem is to recognize that substringy D-brane
physics is not defined by closed string theory or its supergravity limit
but is instead defined by open string theory.  Indeed, very general
world-sheet arguments given in detail at leading order and outlined at
all orders in \DKPS\ show that the leading terms in \stringact\ which survive
the limit \relations\ are described by the theory of the lightest
open strings.
The mass of the lightest open string state
stretching between two D-branes at finite distance $d$ is given by
\eqn\openmass{
 {\widetilde{d} \wm_s^2} = d R m_p^3
}
and thus remains finite, if we do not take the
$R\rightarrow\infty$ limit of \BFSS.  On the other hand, the gap to
the first excited open string state goes as
$\sqrt{(d R m_p^3)^2  + \wm_s^2}-d R m_p^3 \sim \wm_s$
at finite $R$ and hence diverges.

Thus $\wM$ theory on very general grounds is an open string theory,
and because of this many conventional expectations for M theory
are not manifest in the matrix description.
Even the $\sqrt{g}R$ term in \naiveact\ cannot be taken for granted,
because other terms appear to dominate it in the limit.
An example of this is the case of the $\BC^3/\Gamma$ orbifold
target space \dg, where it is argued that the gauge theory construction
does not naturally produce a Ricci flat metric.

A feature of $\wM$ theory not addressed in \DKPS\ is the fact that
the overall size of a compact target space $\CM$ will also shrink to zero
in the limit \relations.  In general this leads to additional light
states, both open strings winding about $\CM$ and D-branes wrapping cycles
in $\CM$.  In small radius limits, general expectations from superstring
duality and much work on toroidal compactification show
that the wrapped D-brane states can be as important as or dominate the
open string states.

On the other hand, we will now argue that for a four dimensional
compact target $\CM$, in the infinite volume limit (in the original
M theory sense), these states are unimportant.  In conventional treatments
of gravity (in space-time dimension greater than three) this limit is
simple to take (there are no significant IR divergences), but given that
gravity is being derived in this framework we must check this point
explicitly.

We assume that $\CM$ is smooth with a volume $V\equiv L^4$
and characteristic curvature length
$l_{C}$.  It may also have non-trivial $p$-cycles with volumes $V_p$;
generically $V_p \sim l_{C}^p$.

The BPS states associated with compactness are
wrapped D$p$-branes,
with mass of order
$$ {\widetilde{V}_p \over \wg_s \wl_s^{p+1}}
  = \wg_s^{(p-4)/3} V_p R m_p^{p+2}, $$
which is infinite for $p < 4$ and $g_s \rightarrow 0$.
The D$4$-brane wrapping on $\CM$ itself has
mass of order $L^4 R m_p^6$ and winding strings have masses
$LR m_p^3$.
These states become arbitrarily massive in the infinite volume
limit in M theory, even though the $\wM$ theory is compactified on
a manifold shrinking to zero size.
Thus we justify the reduction to quantum mechanics in this case.
For D$0$ branes on a resolved ALE space, the gauge theory of \dm\
would describe such a system.

This would break down for compactification
on a six-dimensional manifold, as noted in \refs{\sen,\sei},
and it is not clear whether the large volume limit has associated
subtleties.

\newsec{$\wM$ theory as a naive limit of string theory}

In this section we follow section 2 of \dos\ and
make the assumption that the $\wM$ theory
action should be exactly given by the result of the procedure
of section 2.
We review the computation of the leading supergravity interaction,
a non-BPS result, and show that it does not agree with classical expectations.
We compute the $v^4$ term in the scattering amplitude
of two gravitons on $R^{5,1} \times {\cal M}$. 
The gravitons are fixed at points in ${\cal M}$ 
 but move on $R^{5,1}$ with
relative velocity $v$ and impact parameter $b$.

We first comment that for the result to have any chance to agree
with supergravity, the mass of the lightest open strings must be
equal to $\wsig(x_1,x_2) \tilde{m}_s^2 = \sigma(x_1,x_2) R m_p^3$, where
$\sigma(x_1,x_2)$ is the geodesic distance between the two points
$x_1$ and $x_2$ on ${\cal M}$.
As found in the examples in \dos\ and from the discussion in
\refs{\dcurve,\dko,\strings}, this is a highly non-trivial constraint
on the non-linear sigma model,
and it is not at all obvious from the weak coupling string theory
point of view that it should be satisfied. It should be noted that the
sigma-model on the string worldsheet is strongly coupled in the regime
$\widetilde{\sigma}, \wl_C \ll \wl_s$.
We have no results for or against this point in the present problem,
so in the following we will make the assumption that it is satisfied,
but it could equally well be false, leading to immediate disagreement
with the classical supergravity limit of M theory.

The $v^4$ term at the open string one-loop on K3 has the
following general structure \dos.
\eqn\oneloop{ {\cal A}_{v^4} \simeq \int_0^\infty dt~ t^{5/2} {\rm exp}
 \Big(- t  (\widetilde{b}^2 +
 \wsig^2(x_1,x_2)) \widetilde{m}_s^2 \Big) g(t). }
Here
\eqn\reps{ g(t) = \sum_{i=0}^\infty C_i e^{-h_i t} }
is related to the multiplicities of the $N=4$ superconformal algebra
representations on the open string worldsheet, with
$h_0=0$ corresponds to the lightest state of the open string. 
We can then perform the integral and find, using \relations,
\eqn\onelooptwo{ {\cal A}_{v^4} \propto
  \sum_i { C_i \over (b^2 + \sigma^2(x_1,x_2) +
\wg_s^{-2/3} l_p^2 h_i)^{7/2}}.}

We now see that in
the $\wg_s \rightarrow 0$ limit, the contribution from the massive
excitations disappears provided the gap
${\rm min}_{i \geq 1}(h_i-h_0)$ of CFT spectrum 
remains finite in the limit.  This is not something that
should be taken for granted as the gap may disappear when
CFT becomes singular \refs{\bcov,\shenker,\ov}. In our case, however,
it is easy to see that this is
generically true even in the substringy regime $\widetilde{\sigma},
\widetilde{l}_C \ll \widetilde{l}_s$. First of all, in the orbifold
limit with non-zero theta-angle, we can compute the spectrum explicitly and
show that the gap is finite. We can resolve the singularity by
turning on the marginal perturbation of CFT (corresponding to adding
the Fayet-Illiopoulos term to the D$0$ brane action \dm), and
the gap does not disappear under the small perturbation. 
Therefore the massive excitations decouple in the
limit $\tilde{g}_s \rightarrow 0$, confirming the claim in section 2. 

In this limit, we find
\eqn\onelooplimit{  {\cal A}_{v^4} \rightarrow {C_0 \over
(b^2 + \sigma^2(x_1,x_2) )^{7/2}}}
We can compare this with the supergravity prediction \dos. 
By keeping $x_1$ and $x_2$ sufficiently far away from the orbifold
point, we can set the distance between the points $\sigma(x_1,x_2)$ 
to be smaller than
the curvature radius $l_C$ measured near these points.  
We can then use the short distance expansion of
the gravition scattering amplitude and compare it with \onelooplimit. 
We find that subleading terms in the expansion are missing here.
As pointed out in \dos,  the
discrepancy remains even if there are corrections to the mass of the
lightest open string state, as long as the massive excitations
decouple and we are left with the gauge theory degrees of freedom.

Now at large $N$ it is possible that the subleading correction to
\onelooplimit\  will appear as bound
state effects, but for $N_1=N_2=1$ this is not possible.

We note that the same problem can be studied in other curved
backgrounds, and the general properties of gauge theory and supergravity
are so different as to make universal agreement quite unexpected.
As an example,
the dynamics of D$0$-branes in a five-brane background can be studied,
and in this case even the relation
$m_W = \wsig(x_1,x_2) \widetilde{m}_s^2$ appears to fail \dd.

\newsec{Possible resolutions}

The computation we described is valid at finite $N$ and thus the
result should be compared with DLCQ supergravity.
Could this interaction be different?

The first guess one might try is that supergravity loop effects are
important.  On physical grounds this requires some energy or momentum
in the problem comparable to $m_p$, while we find disagreement even in
the case of all length scales large compared to $l_p$.

The only new feature of DLCQ is the quantization $p_-=N/R$ and from
this point of view $N$ small is advantageous.  We should take
the conventional parameter $R \gg m_p^{-1}$.
We also require $m_p \gg p_+=p_\perp^2/2p_-$ and here
$N$ small means that we must have $p_\perp^2 \ll m_p^2$ to have both
$p_+ \ll m_p$ and $p_- \ll m_p$.
The resulting uncertainty relation
$\Delta x \sim 1/p$ leads to no further restrictions.
We conclude that the usual energy suppression of loop effects in gravity
is still valid in DLCQ.

Even on the classical level,
DLCQ supergravity has not been much studied in the past, with some
exceptions \gorschw, and so it
is conceivable that there are other subtleties relevant for this problem.

\medskip

We finally come to our suggested resolution.
The assumption made in section 3 that Matrix theory can be obtained
by a naive application of the limiting procedure of
\refs{\sen,\sei}\ leads to contradictions with the expected physics
of M theory.

Does this mean that Matrix theory is wrong ?  Not necessarily,
but it appears to mean that
the boost to the infinite momentum frame is complicated.

Evidence for this idea can be found in the study of the analogous
case of light-front quantization of gauge theory.
(Some representative recent work can be found in \lcrefs.)
This has been much studied over the years, especially as a
non-perturbative approach to QCD, and has numerous advantages over
the equal-time canonical approach.  For example, in a theory without
massless particles, the perturbative vacuum is the Fock vacuum -- the
complicated problem of solving for the vacuum state is eliminated.

These advantages do not come for free and the loss of Lorentz invariance
means that the problem of renormalization is much more complicated.
In particular, modes with longitudinal momentum $p_-=0$ must be
integrated out to derive such a canonical formulation, and this
leads to non-local divergences and counterterms which can only be
determined by requiring Lorentz invariance in the resulting physics.

We can compare this with the discussion in \strings.
There it was argued based on the results of \refs{\dos,\dg}\ and
\DKPS\ that although one can use D$0$-brane theories as definitions of
Matrix theory, reproducing expectations from eleven-dimensional supergravity
would require using a Lagrangian with
renormalized parameters and possible additional interactions.

What consistency conditions determine the renormalized Lagrangian?
We cannot impose eleven-dimensional Lorentz invariance in a general
curved background.
On the other hand, it is known that Lorentz invariance in flat space,
the presence of spin two degrees of freedom,
and very general consistency conditions lead to the Einstein action
for the low energy physics of gravity \Weinberg.
Adding supersymmetry to this, we can ask as our substitute for
Lorentz invariance that supergravity amplitudes be correctly reproduced
starting with a non-singular Lagrangian.
It was shown in \refs{\strings,\dko}\ that this leads to very non-trivial
constraints on the Lagrangian, which do admit solutions.
(It should be pointed out that the conditions studied so far do not
completely determine the matrix model action \dko.)

Thus we suggest as the resolution of the paradox the idea that the boost
to the infinite momentum frame renormalizes the Lagrangian in a way which
will lead to correct eleven-dimensional physics.

Another consideration which is compatible with this idea is to study
the inverse process of deriving a weak string coupling D$0$-brane
Lagrangian from Matrix theory.  The way to do this is outlined in \dvv;
they show that the $1+1$-dimensional gauge theory describing Matrix
theory on $S^1$ renormalizes to a world-sheet theory of an arbitrary
light-cone gauge strings, with the correct interaction.  States with
D$0$-branes are described as sectors with electric flux in the gauge
theory, and they argue that in the renormalized theory strings will end
on D$0$-branes in the appropriate way.

One could follow the same approach for compactification on $S^1\times\CM$
to derive the corresponding string theory compactification and
D$0$-brane Lagrangian.  The non-trivial renormalization in this process
means that we can expect to get sensible world-sheet Lagrangians from
a wide variety of starting points; there is no reason at all from this
point of view why the starting Matrix theory Lagrangian and resulting
D$0$-brane Lagrangian should be the same.

An interesting counterpart to the idea comes from considering
further compactification on $T^n\times\CM$.  For $n>1$ and non-compact
$\CM$, this would lead to an $n+1$-dimensional
supersymmetric non-linear sigma model,
which is generally believed to be non-renormalizable.

Now that we believe that sensible interacting field theories in dimensions
greater than four can exist, it seems premature to assume
that such supersymmetric non-linear sigma models cannot be renormalizable.
It might be that renormalizable models exist, and that this constraint will
provide a more effective way of determining the correct Lagrangians.

To add to the mystery surrounding this question, however, we must note
that for $\CM_4$ compact and $n>1$, as we mentioned,
it could be that the large volume limit is subtle.
This comment might also be relevant to the contradiction that supergravity
predicts that six dimensional
Calabi-Yau manifolds with $R l_p^2 \ll 1$ will be Ricci flat,
while weak coupling string theory with $R l_p^2 \ll 1 \ll R l_s^2 $
predicts that D$0$-branes do not see a Ricci flat metric \refs{\dg,\ggv},
which otherwise would be direct evidence for our proposal.

However, these subtleties do not enter for the case of K3 compactification,
and our overall conclusion is that the Matrix theory Lagrangian must be
different from the weak string coupling Lagrangian in order to reproduce
the physics of M theory.

A priori the relation between these two Lagrangians
is likely to be quite complicated, and this
is the meaning of the title of the paper.  Of course
the Matrix theory has already provided
simpler arguments for duality as well as much insight; however
these issues will have to be addressed to get non-BPS results.

\bigskip
\noindent
{\bf Note added}
\medskip

After the completion of this work we received a paper \dr, where they
found a discrepancy between the three graviton scattering amplitude
calculated from the matrix theory and the supergravity prediction. As
we are studying the two graviton scattering in a weak gravity
background, there may be a relation between the results.  A strong test
of the idea described here will be to find a modified maximally
supersymmetric Lagrangian which fixes this problem.

\bigskip
\noindent
{\bf Acknowledgements}
\medskip

We would like to thank Tom Banks, Kiyoshi Higashijima, Ashoke Sen, Koichi
Yamawaki and, in particular, Steve Shenker for discussion.
H.O is supported in part by NSF grant PHY-951497 and DOE grant
DE-AC03-76SF00098.

\listrefs\end